\documentclass[a4paper,11pt]{article}
\usepackage{pos}

\title{Reducing model uncertainties using proton-oxygen collisions with proton/neutron tagging at the LHC}
 \ShortTitle{Reducing uncertanties with proton/neutron tagging at the LHC}

\author*[a,b]{Michael Pitt}

\affiliation[a]{Ben-Gurion University of the Negev, Department of Physics,\\
  Beer-Sheva, Israel}

\affiliation[b]{The University of Kansas, Department of Physics,\\
Lawrence, USA}

\emailAdd{michael.pitt@cern.ch}

\abstract{

A short run of proton-oxygen and oxygen-oxygen collisions is planned to take place at the Large Hadron Collider during LHC Run 3. The primary goal of this run is to improve the modeling of Cosmic-Ray interactions and to reduce the uncertainties associated with proton-Air cross-sections. While the inelastic cross-section will be measured directly, an array of very forward proton and neutron detectors introduced by the ATLAS and CMS experiments can allow going beyond the current physics research proposal, providing a unique opportunity to study elastic and diffractive interactions in pO collisions at the center of mass energies above TeV. This article presents the possible impact of proton and neutron tagging on the measurement of the elastic and diffractive components, as well as discusses the prospects of measuring the decay products of oxygen ions.
}

\ConferenceLogo{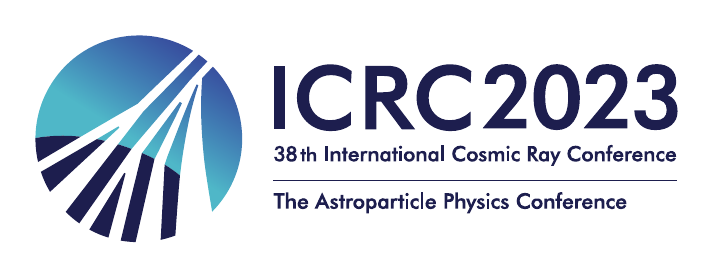}

\FullConference{%
38th International Cosmic Ray Conference (ICRC2023)\\
  26 July - 3 August, 2023\\
  Nagoya, Japan}

\begin{document}
\maketitle

\section{Introduction}

\begin{figure}[htb]
\centering
\includegraphics[width=0.4\textwidth]{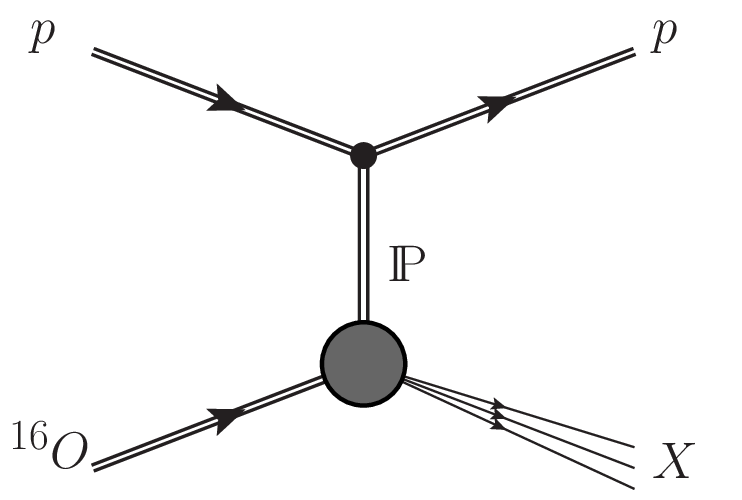}
\includegraphics[width=0.4\textwidth]{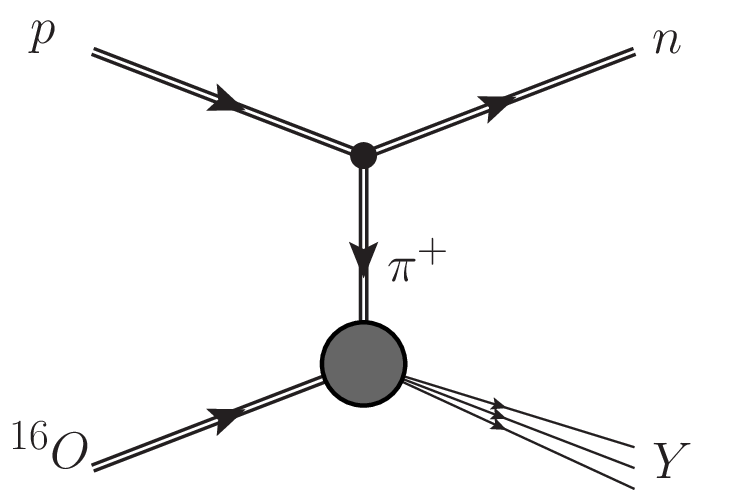}
\caption{Schematic diagrams of $pO$ collisions with a pomeron exchange (left) or a pion exchange (right), resulting in an emerging forward proton or neutron, respectively.\label{fit:process}}
\end{figure}

Cosmic rays (CRs) span a wide range of energies, extending up to 10$^{21}$ eV. The nature and origin of ultra-high-energy CRs (with energies above 10$^{18}$ eV) is a subject of extensive study. The energy and identity of such CRs can be studied through the extended air showers produced when CRs collide with the upper atmosphere on earth. Determining their mass and energy would help clarify the origin of the most energetic particle in the Universe. The estimation of these parameters hinges upon measuring and simulating the maxima air-shower profiles ($X_\text{MAX}$). The modelling of the air-shower profiles, is done using hadronic Monte Carlo (MC) simulations \cite{mc1,mc2}. Some MC event generators are tuned based on the measured inelastic cross-sections in proton-proton, proton-lead, or lead-lead interactions at the Large Hadron Collider (LHC), the diffractive component remains weakly constrained. Notably, there are substantial discrepancies between the experimental data and prediction of MC simulations in cases like proton-lead collisions \cite{CMS:2023lfr}. To improve the modeling of hadronic interaction and reduce model uncertainties associated with proton-Air cross-section modeling, a short run of proton-oxygen ($pO$) collisions during LHC Run 3 has been proposed \cite{Bruce:2021hjk}. While the inelastic cross-section will be measured directly \cite{Brewer:2021kiv}, the elastic and diffractive interactions in pO collisions will remain unexplored. Tagging forward protons in $pO\to pX$, or forward neutrons in $pO\to nY$ interactions will provide a unique opportunity to study those components of the total proton-oxygen cross-section, and are the main subject of this article. Schematic diagrams of processes of interest are illustrated in figure \ref{fit:process}.

\section{Proton and neutron tagging at the LHC}

\begin{figure}[!htb]
\centering
\includegraphics[width=1.0\textwidth]{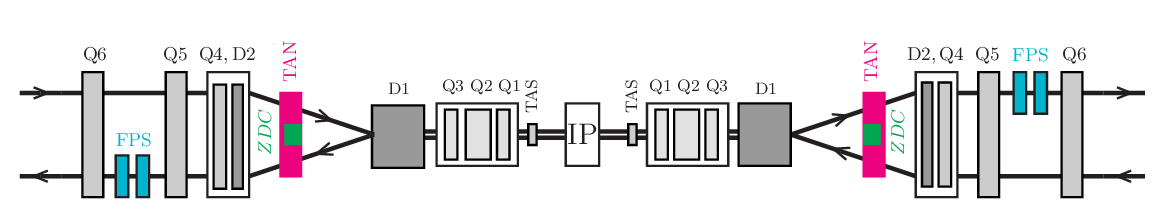}
\caption{Schematic layout of the insersion region between the interaction point and quardupole magnet Q6.  to the Q6.\label{fig:layout}}
\end{figure}

Forward neutron and proton detectors have significantly expanded the scope of the Heavy Ion and proton-proton physics programs of the ATLAS and CMS experiments. The arrangement of the detector devices along the LHC beamline, on both sides of the Interaction Point (IP), with insertion magnets and absorbers is schematically illustrated in figure \ref{fig:layout}.

The first inner triplet of quadrupole magnets (Q1-3) is shielded from the high-energy charged and neutral particles produced at the IP by the target absorber (TAS). The neutral beam absorber (TAN) positioned between separation and recombination dipole magnets (D1, D2) protects machine components against neutral particles emerging from the IP and is used to host the neutron detectors. Following the large aperture Q4 quadrupole magnet, the proton detectors are located in the region between two quadrupole magnets, Q5 and Q6.

\subsection{Forward Proton Spectrometer (FPS)}

The Forward Proton Spectrometers (FPS), introduced during LHC Run 2 by the ATLAS and CMS collaborations, consist of near beam detectors located at about 200~meters from the IP. Operated during the standard high luminosity LHC runs, these spectrometers are primarily used to study the central exclusive production processes in proton-proton collisions. Both experimental apparatus, the ATLAS Forward Proton detector (AFP) \cite{Adamczyk:2015cjy} and CMS-TOTEM Precision Proton Spectrometer (CT-PPS) \cite{CMS:2014sdw} have been seamlessly integrated during the standard LHC runs and delivered a broad range of physics results.

Besides probing the central exclusive production processes, proton tagging offers a distinctive avenue for investigating the elastic and diffractive components when operated at relatively low proton-proton collision rates. The range of kinematic acceptance for protons is contingent upon the magnetic field of the LHC. Protons interacting diffractively lose a fraction of their momentum (denoted by $\xi=\Delta p/p$) deflected away from the beam center. The LHC magnetic field determines proton displacement, and the kinematics are computed by inverting the signal-pass proton transport matrix defined by the optical functions that describe the proton transport in the vicinity of the so-called central orbit \cite{Nemes:2017icw}. FPS acceptance with LHC optics used in the standard runs LHC typically ranges from 1.5\% to 15\%. This range of proton acceptance is anticipated to hold for the forthcoming $pO$ collisions.

To incorporate the FPS into the proton-oxygen run, a detector alignment process must be performed. The alignment can be achieved when the FPS approaches the beam up to a few beam sigma until splashes are detected in the radiation monitors, determining the distance to the beam center. Further refinement of the alignment process requires operating LHC for several hours at very low beam intensity, as outlined in \cite{Kaspar:2017gwd,cms:propog}.

Following the complete alignment sequence, one can determine the charge particle displacement from the beam center and follow the optical functions to derive the hadron's transverse momentum and the momentum loss ($\xi$). Two distinct measurements can be performed using proton tagging during the $pO$ runs and will be discussed subsequently:

\begin{itemize}
\item The measurement of the diffractive component of the total $pO$ cross-section by tagging the intact protons in $pO$ collisions.
\item Determination of light ion production rates (form Z=1 to Z=8) in $pO$ and $OO$ collisions by tagging outgoing light ions in FPS. 
\end{itemize} 

\subsection{Zero Degree Calorimeter (ZDC)}

The Zero Degree Calorimeter (ZDC) is a specialized detector placed at a zero-degree angle concerning the beamline, used to detect forward neutral particles produced in $AA$ and $pA$ collisions, primarily spectators arising from ion collisions. In both ATLAS and CMS interaction regions of the LHC, the ZDC is installed in a dedicated slot inside the neutral beam absorbers (TAN), located at a distance of 140~meters from the IP.

The ZDC plays a crucial role in detecting forward neutrons and photons with $|\eta|>8.5$ during $pp$, $pA$, and $AA$ collisions. In $pp$ collisions, the ZDC detectors can operate at instantaneous luminosities well below $10^{33}\text{cm}^{-2}\text{s}^{-1}$. The ZDC comprises an electromagnetic module, approximately 30 radiation lengths long, and three hadronic modules, each about 1.15 interaction lengths long. Notably, vetoing events involving spectators is a core strategy for tagging ultra-peripheral collisions. The design of the ZDC enables the determination of kinematics and production cross-sections for forward-going neutral pions, kaons, and eta mesons. The ZDC provides crucial data on light meson production from protons at LHC energies that cannot be obtained through other means.

\section{New constrains on MC hadronic models}

Hadronic MC simulations are tuned using available experimental data and are applied across various research domains. One notable example involves the study of ultra-high-energy CRs through the extended air showers produced when CRs collide with the upper atmosphere on Earth. Determining their mass and energy stands to elucidate the origins of these particles, contingent upon a thorough analysis of the measured and simulated maxima air-shower profiles ($X_\text{MAX}$). Large uncertainties stemming from different modeling of hadronic interactions weaken the constraints of cosmic ray mass composition \cite{ref1}. For example, the EPOS-LHC - MC event generator for minimum bias hadronic interactions is tuned to the measured inelastic cross-sections in proton-proton, proton-lead, or lead-lead interactions at the LHC. Therefore the measured diffractive signatures in proton-lead collisions differ from the prediction of the MC event generators \cite{CMS:2023lfr}.

\begin{figure}[htb!]
\centering
\includegraphics[width=0.45\textwidth]{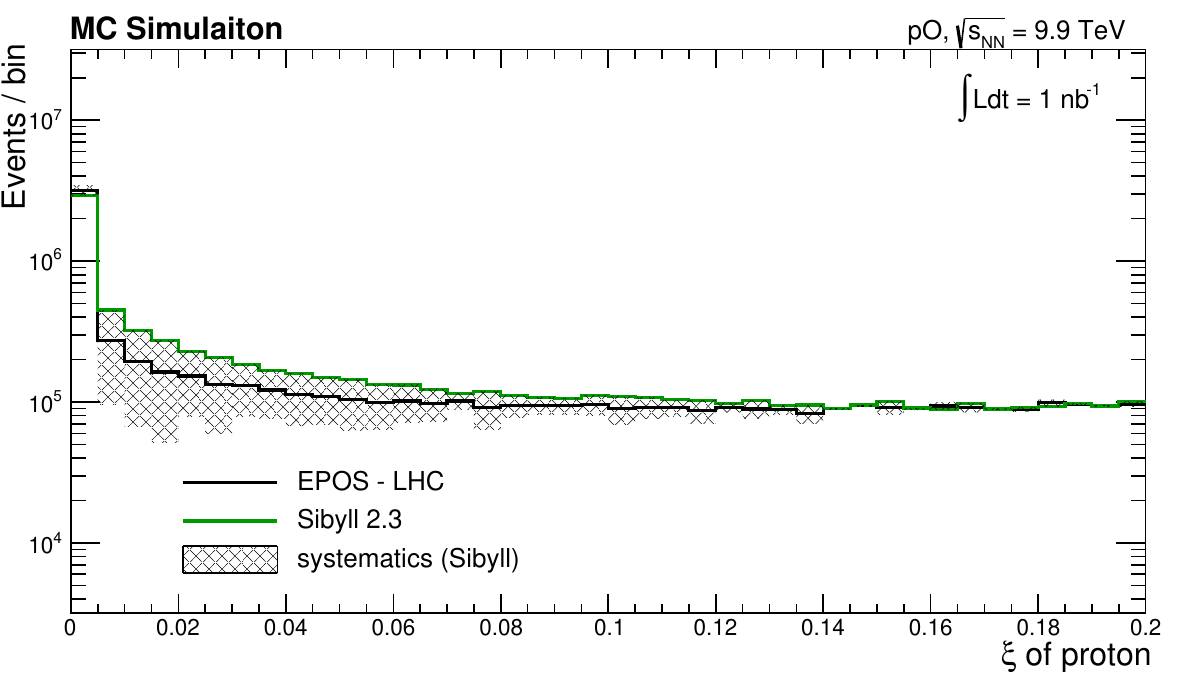} 
\includegraphics[width=0.45\textwidth]{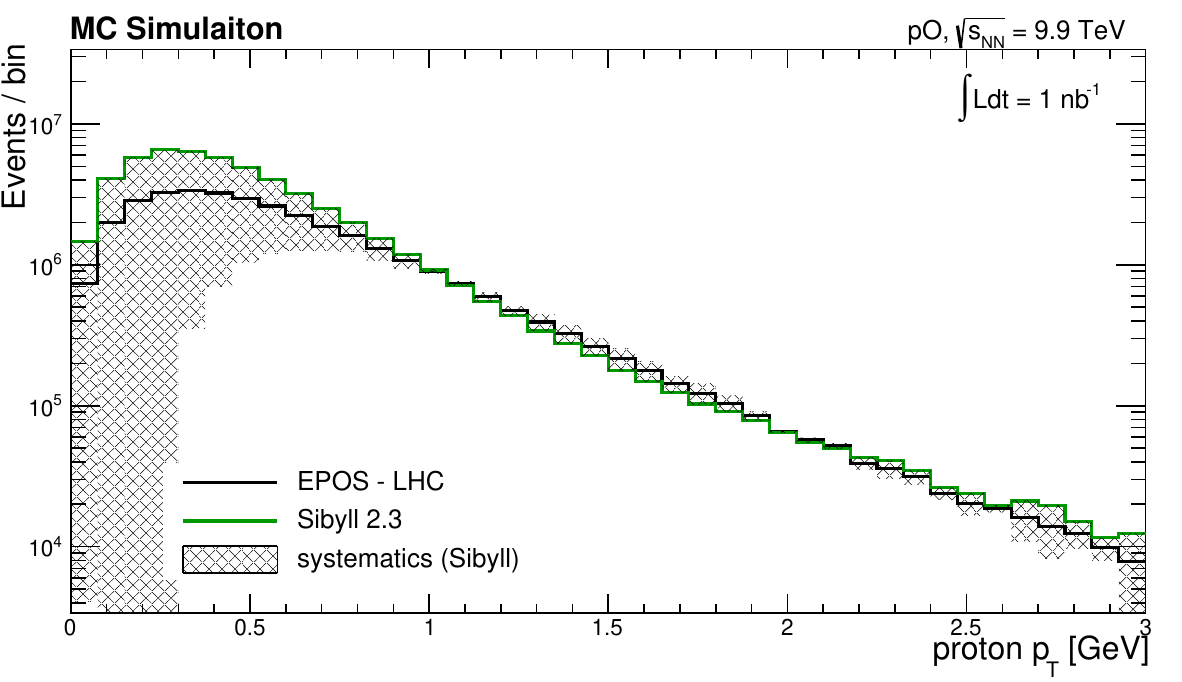}
\caption{Differential cross section $d\sigma/d\xi$ (left) and $d\sigma/dp_{T}$ (right) for $pO$ collisions assuming integrated luminosity of $L_\text{int}=1nb^{-1}$ at center-of-mass energy per nucleon pair of $\sqrt{S_\text{NN}}=9.9$ TeV, obtained using the EPOS-LHC and Sibyll 2.3 MC event generators The dashed area represent the model uncertainty derived from the comparison between the two models.\label{Fig:proton_compare}}
\end{figure}

During the proton-oxygen run, proton kinematics in diffractive and elastic interactions can be determined by operating the FPS downstream of the proton beam. Utilizing proton tagging enables the exploring colorless interactions in pO collisions (including elastic, diffractive, and pion exchange processes), which represent about 20\% of the total cross-section. As the FPS proton momentum loss acceptance ranges from 1.5\% to 15\%, these detectors can tag a subset of 2-4\% of all $pO$ events. The FPS detectors exhibit the capability to precisely measure proton kinematics, allowing refining predictions made by various hadronic models. To illustrate, a comparison highlighting the disparity in predicted proton kinematics between two Monte Carlo event generators is depicted in figure~\ref{Fig:proton_compare}.

\begin{figure}[htb!]
\centering
\includegraphics[width=0.6\textwidth]{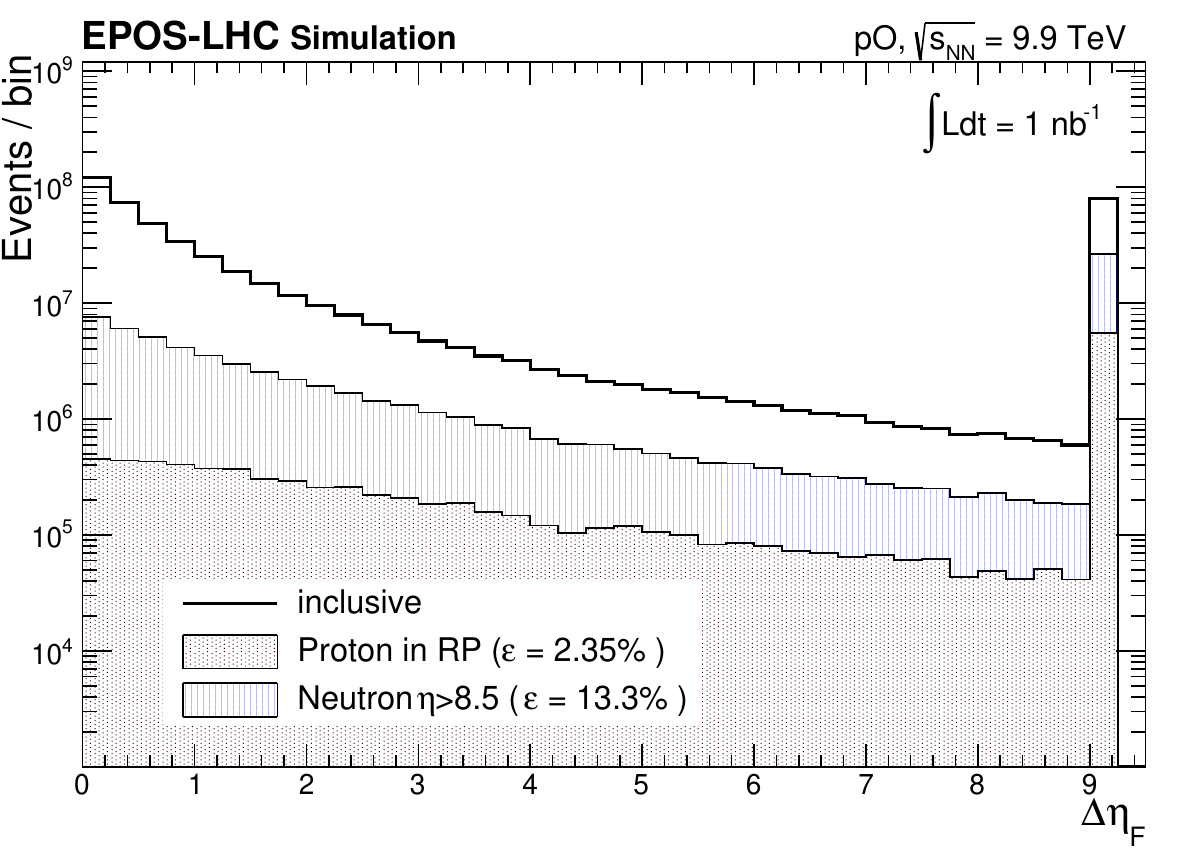}
\caption{Contribution from $pO\to pX$ and $pO\to nX$ interactions with a proton within the FPS acceptance or a neutron within the ZDC acceptance. The peak at $\Delta\eta_\text{F}$ distribution corresponding to zero bias events (no particles with energy above 1 GeV are detected within the pseudorapidity range of $|\Delta\eta| < 4.5$).\label{Fig:fig3}}
\end{figure}

Elastic and diffractive contributions manifested by large gaps in the rapidity distribution of final-state particles, defined as $\Delta\eta_F$. While the probability of finding a continuous rapidity region $\Delta\eta_F$ free of particles is suppressed exponentially in non-diffractive inelastic events, discriminating different topologies of colorless interaction (pomeron or pion exchange) is only achievable through proton and neutron tagging. Figure~\ref{Fig:fig3} illustrates the contribution from diffractive processes where a proton is measured by FPS or a neutron measured by the ZDC as a function of the $\Delta\eta_\text{F}$ spectra.

\section{Light ion production rates}

Interestingly, the magnetic fields of the ion beam exhibit similarities to those of the proton beam in terms of charged particle transport, implying the potential for detecting oxygen ions within the FPS with a similar momentum loss range of 1.5\% to 15\%.

\begin{figure}[htb!]
\centering
\includegraphics[width=1\textwidth]{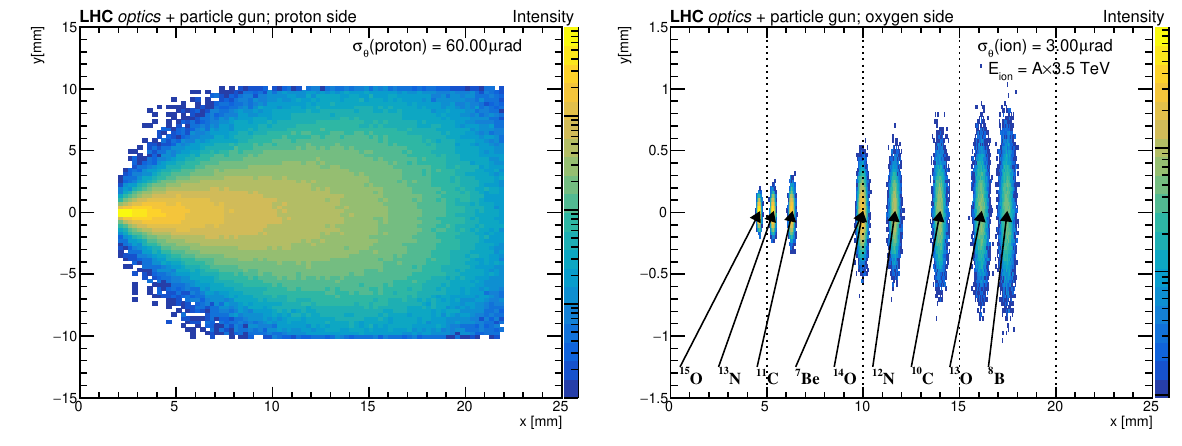}
\caption{The simulated hit pattern of charged particles in the FPS detector sensors located downstream the proton (left) and oxygen (right) beamlines. In this scenario, proton displacement is governed by the momentum loss of protons ($\xi$), while ion displacement is determined by the $A/Z$ ratio. Several prominent hot spots on the right plot indicates different ion species.\label{Fig:fig6}}
\end{figure}

During hard scattering, oxygen ions will disintegrate, yielding light ions alongside protons and neutrons resulting from the nuclear break up of oxygen ions. No measurements of the abundance of light ions from heavy ion collisions at TeV scale energies exist. The majority of light ions originating from ion disintegration are expected to possess a momentum around $0.5\times E_p \times A$, where A is the mass number of the light ion, and $E_p$ represents the energy required to maintain a proton in a stable orbit. The light ions with different $A/Z$ ratios will behave as the nominal beam particle with momentum loss of $\xi = 1 - 0.5\times A/Z$. Spectator protons will carry half of the beam energy ($\xi=0.5$) and escape detection. Only isotopes with $1.7 < A/Z < 1.97$ could be measured by the FPS detectors. In such a scenario, FPS will serve as a mass spectrometer. A hit pattern from simulated protons and ions propagated using the LHC transport matrix to the location of the FPS detectors is shown in figure \ref{Fig:fig6}.

\section{Conclusions}

Including forward proton and neutron detectors during the LHC oxygen run presents an exceptional opportunity for the physics research program. This program aims to significantly constrain diffractive and elastic interactions in proton-ion collisions with high precision. Additionally, it seeks to measure the elastic component of proton-ion interactions for the first time. Furthermore, by conducting further investigations into ion disintegration and production rates, it may be possible to achieve ground-breaking measurements in this domain. Successful implementation of this research program could serve as a stepping stone for future measurements utilizing both FPS and ZDC in heavy ion runs at the LHC.

%
%
%

\end{document}